\documentstyle[12pt]{article}

\newcommand{\nc}{\newcommand}
\def\frac#1#2{{\textstyle {#1 \over #2}}}

\nc{\beq}{\begin{equation}}
\nc{\eeq}{\end{equation}}
\nc{\beqa}{\begin{eqnarray}}
\nc{\eeqa}{\end{eqnarray}}
\nc{\lsim}{\begin{array}{c}\,\sim\vspace{-21pt}\\< \end{array}}
\nc{\gsim}{\begin{array}{c}\sim\vspace{-21pt}\\> \end{array}}

\def\&{and}

\def\nc#1#2#3{           {\it Nuovo Cim.  }{\bf #1}, #2 (19#3)}

\begin{document}

\begin{titlepage}

\begin{center}
     \hfill       BUHEP-98-08\\
\vskip .5 in
{\Large \bf Duality in Non-Supersymmetric MQCD}
\vskip .3 in


{{ Nick Evans }}

   \vskip 0.3 cm
   {\it Department of Physics,
        Boston University,
        Boston, MA 02215}\\
\end{center}

\vskip .5 in
\begin{abstract}
\noindent
The curves that describe the M-theoretic extension of type IIA string
configurations with non-supersymmetric field theories on their surface
exhibit a duality map. The map suggests a continued link between a 
$SU(N)$ gauge theory with $F$ flavours and an $SU(F-N)$ gauge theory with $F$
flavours (the duality of supersymmetric QCD) 
even when the gaugino mass is taken to infinity. Within the
context of the field theory such a duality only continues to make sense if
the scalar fields remain light. We discuss the difficulties of
decoupling the scalars within this framework. 
\end{abstract}
\end{titlepage}

\input epsf
\newwrite\ffile\global\newcount\figno \global\figno=1
\def\writedefs{\immediate\openout\lfile=labeldefs.tmp \def\writedef##1{%
\immediate\write\lfile{\string\def\string##1\rightbracket}}}
\def\writestoppt{}\def\writedef#1{}

\def\figin{\epsfcheck\figin}\def\figins{\epsfcheck\figins}
\def\epsfcheck{\ifx\epsfbox\UnDeFiNeD
\message{(NO epsf.tex, FIGURES WILL BE IGNORED)}
\gdef\figin##1{\vskip2in}\gdef\figins##1{\hskip.5in}
\else\message{(FIGURES WILL BE INCLUDED)}%
\gdef\figin##1{##1}\gdef\figins##1{##1}\fi}

\def\figinsert{}
\def\ifig#1#2#3{\xdef#1{fig.~\the\figno}
\writedef{#1\leftbracket fig.\noexpand~\the\figno}%
\figinsert\figin{\centerline{#3}}\medskip\centerline{\vbox{\baselineskip12pt
\advance\hsize by -1truein\center\footnotesize{  Fig.~\the\figno.} #2}}
\bigskip\endinsert\global\advance\figno by1}
\def\footnotefont{}\def\endinsert{}

\renewcommand{\thepage}{\arabic{page}}
\setcounter{page}{1}

\section{Introduction}

The improved understanding of supersymmetric QCD over recent years has
revealed many new aspects of gauge dynamics. Of particular interest
from the study of N=1 SQCD have come examples of gauge theories that
give rise to massless composites and exhibit a non-abelian
electromagnetic duality \cite{dual}. 
Explicitly, N=1 $SU(N)$ SQCD with $F$ matter
flavours, $Q, \tilde{Q}$,
 is dual to an $SU(F-N)$ gauge theory with $F$ flavours $q, \tilde{q}$, 
a meson superfield, $M$,
 transforming as $(F, \bar{F})$ under the flavour symmetry,
and a superpotential $Mq \tilde{q}$. Whether this sort of gauge dynamics
is special to supersymmetric theories remains unclear. Soft
supersymmetry breaking parameters may be introduced through the vevs of
higher component fields of spurion superfields (the couplings and
parameters of the theory) \cite{soft1}-\cite{N=2soft}. 
The symmetries of the model are not though
sufficient to allow a map of soft breakings in the electric variables to
those in the magnetic variables except in a few special cases \cite{soft2}. 
Scalar masses result from Kahler terms that are non-holomorphic and so
in general it is unclear whether the dual theory reacts to supersymmetry
breaking in the alternative variables by a mass or higgs branch
\cite{soft3}. Whether
the duality persists with massless states, the theory  develops a mass
gap, or whether the dual variables of SQCD are no longer the relevant 
degrees of freedom as
supersymemtry is broken is therefore unclear. Recently \cite{soft4} 
has suggested methods to overcome these issues for perturbing soft breakings.

A large literature has grown up on engineering supersymmetric
field theories using D-brane constructions in type IIA string
theory (a comprehensive review and list of references is to be found in
\cite{brane}). 
The essential ingredient is that the massless string modes of strings
ending on the surfaces of D-branes correspond to gauge fields living on
the D-branes surface plus superpartners. Whilst a perturbative identification between the string states in the
type IIA theory and the UV field theory may be made the IIA picture
provides no information about the strongly coupled 
IR dynamics of the theory. Such information would correspond to short
distance structure of the branes but due to the strongly coupled nature
of the core of NS5 branes the precise structure of an NS5 D4 junction is
unclear. To proceed one must move to M-theory \cite{witten1} 
(type IIA string theory
with coupling $g_s$ is the 11 dimensional M-theory compactified on a
circle of radius, $R \sim g_s$). In M-theory NS5 branes
and D4 branes become aspects of a single M5 brane wrapped in places on
the compact dimension. The junctions between these objects may thus be
smoothly described by a minimal area 
embedding of the M5 brane. Increasing the M-theory compactification
radius from zero allows the study of the string theory with increased
coupling at the string scale. It is therefore possible to make a strong
coupling expansion to the field theory, that is, smoothly deforming the
field theory of interest to a theory with the same global symmetries and
parameters but which is fundamentally a theory of strongly interacting
strings in the $R \rightarrow \infty$ limit. For intermediate $R$ the
theory has Kaluza Klein states in addition to those of the field theory. 
We hope that by making this transition between smoothly related 
theories there is no phase transition and that the
two theories ly in the same universality class. This technique has been
used to derive the existence of gaugino condensation in N=1 super Yang
Mills theory \cite{witten2} and duality \cite{branedual} \cite{Mdual} 
in the theories with flavour amongst other
results (see \cite{brane}). 
For these supersymmetric theories the holomorphic and BPS
properties of states in the theory contribute to making the motion to
strong coupling smooth.

In \cite{nsbrane}-\cite{barbon2} 
the construction of non-supersymmetric theories on D-brane
surfaces has been discussed. The branes are rotated relative to their 
positions in the supersymmetric configurations.
In the perturbative type IIA string theory the supersymmetry breaking 
of the branes can be mapped to supersymmetry breaking in the field
theory using the strict constraints of N=2 supersymmetry on how spurions
may enter the field theory \cite{nevans1}\cite{nevans2}. 
M-theory curves for a subset of these configurations with and
without matter fields have been proposed \cite{witten2}\cite{nevans2}. 
The low energy $Z_N$
symmetries of the supersymmetric models are broken by the
supersymmetry breaking terms in the curve and the degeneracy of the $N$
supersymmetric vacua is lifted. The curves display the same
phase structure with changing theta angle as softly broken field
theories \cite{barbon1}\cite{nevans2}. 
For the case with matter fields it has been shown that if
parameters are chosen to decouple the gaugino then the low energy theory
is described by two parameters, the quark mass and a parameter with the
charges of a quark condensate \cite{nevans2}. 

In this paper we show that the non-supersymmetric M-theory curves
exhibit the same duality map as the supersymmetric curves. We discuss
whether non-supersymmetric gauge theories could possess a non-abelian
duality and conclude that it is only consistent if the scalar fields are
remaining relevant to the low energy dynamics. A similar conclusion has
been reached recently in \cite{barbon2}. With the loss of
supersymmetry and its associated holomorphic and BPS properties we must
be careful about treating the brane picture with its extra states
as a proof of the gauge
theory behaviour but the result is suggestive that the SQCD
duality persists in the limit of the gaugino decoupling and fine tuned
light scalars. 

One might have hoped to be able to decouple the scalar fields and learn
about the transition from SQCD to honest QCD without scalar fields from
the brane set up. Unfortunately the available brane deformations do not
seem to include those corresponding to positive masses for all
scalars and hence this decoupling can not be performed at tree level. 
Nevertheless it
is clear that QCD can not possess a description in terms of a flavour
dependent dual gauge group, such as that suggested by the branes,
since QCD has no higgs branch.

In section 2 we review the 4D field theories that may be engineered with
type IIA string theory branes. In section 3 we review the derivation of
duality in N=1 SQCD with branes following \cite{branedual}\cite{Mdual}. 
In section 4 we
demonstrate the duality map holds for non-supersymmetric M-theory brane
configurations and in section 5 discuss the implications for the field
theories on their surfaces.

\section{Type IIA Configurations and Field Theories}

The basic brane configuration from which we start corresponds
to $SU(N)$ N=2 SQCD with $F$ quark matter flavours. It is given from left to
right by the branes

\beq
\begin{tabular}{|c|c|c|c|c|c|c|c|c|}
\hline
 & $\#$ & $R^4$ & $x^4$ & $x^5$ &  $x^6$ & $x^7$ & $x^8$ & $x^9$ \\
\hline 
NS5 & 1 & $-$ & $-$ & $-$  &  $\bullet$ & $\bullet$ & $\bullet$ & $\bullet$ \\
\hline
D4 & $N$ & $-$  & $\bullet$ &  $\bullet$ &  $[-]$ & $\bullet$ & 
                              $\bullet$ & $\bullet$ \\
\hline
NS5$'$ & 1 & $-$ & $-$ & $-$ & $\bullet$ & $\bullet$  &  $\bullet$ & $\bullet$ \\
\hline
D4$'$ & $F$ & $-$  & $\bullet$ &  $\bullet$ & $-$ & $\bullet$ & 
                              $\bullet$ & $\bullet$ \\
\hline
\end{tabular} 
\eeq

$R^4$ is the space $x^0-x^3$. A dash $-$ represents a
direction along a brane's world wolume while a dot $\bullet$ is
transverse. For the special case of the D4-branes' $x^6$ direction,
where the world volume is a finite interval corresponding to their
suspension between two NS5 branes at different values of $x^6$, 
we use the symbol $[-]$. 
The field theory exists on scales much greater 
than the $L_6$ distance between the NS5 branes with  the fourth
space like direction of the D4-branes generating  the couplings of the
gauge groups in the  effective 4D theory. The semi-infinite D4$'$ brane
is responsible for providing the quark flavours.

The $U(1)_R$ and $SU(2)_R$ symmetries of the N=2 field theory are
manifest in the brane picture. They correspond to isometries of the
configuration; an SO(2) in the  $x^4,x^5$ directions and an SO(3) in the
$x^7, x^8, x^9$ directions.

N=2 supersymmetry may be broken in the configuration by rotations of the
branes away from this configuration of maximal symmetry. In the field
theory these breakings correspond to soft breakings of supersymmetry
introduced through the vevs of auxilliary components of
spurion fields (couplings) of the theories
\cite{nevans1}\cite{nevans2}. 
The N=2 supersymmetry is
sufficiently restrictive on how these spurions may enter the field
theory that the correspondence between the brane motions and field
theory parameters can be identified.  The
possible breakings are:

\begin{itemize}
{\item 
The NS5 branes may be rotated
in the  $x^4,x^5,x^7, x^8, x^9$ space (rotations from the N=2
configuration into the $x^6$
direction cause the NS5 branes to cross changing the topology of the
configuration in such a way that it can no longer be easily identified
with a field theory). These rotations correspond to components of the 
spurion fields occuring as  vector fields 
in the prepotential of the N=2 theory as ${\cal F} = (S_1 + i
S_2)A^2$ \cite{N=2soft}. 
The scalar spurion vevs generate the gauge coupling $\tau$.
When we allow the auxilliary fields of the spurions to be
non-zero we obtain the tree level masses

\begin{eqnarray} \label{softUV}
&& \nonumber
-{N_c \over 8  \pi^2} Im \left( (F_1^*+iF_2^*) \psi_A^\alpha \psi_A^\alpha 
+ (F_1 + i F_2)
\lambda^\alpha \lambda^\alpha + i \sqrt{2}
(D_1 + i D_2) \psi_A^\alpha\lambda^\alpha \right) \\
&&
  - {N_c \over 4 \pi^2 Im (s_1+is_2)} \left((|F_1|^2 + D_1^2/2) Im(
a^\alpha)^2
+  (|F_2|^2 + D_2^2/2) Re
(a^\alpha)^2 \right. \\
&& \left. \right. \hspace{4cm} + \left. 
 (F_1 F_2^* + F_1^*F_2 + D_1D_2) Im(a^\alpha) Re(a^\alpha)
\right)\nonumber
\end{eqnarray}
A number of consistency checks support the identification \cite{nevans1}. 
Switching on
any one of the six  independent real supersymmetry breakings in the field
theory leaves the same massless spectrum in the field theory as in the
brane picture when any one of the six independent rotations of the NS5
brane is performed. The field theory and brane configurations possess
the same sub-manifold of N=1 supersymmetric configurations (we
concentrate on these configurations below).}
{\item Supersymmetry may also be broken by forcing the D4 
and D4$'$ branes to lie at
angles to each other.
With the introduction of matter fields in the field theory a single
extra spurion field is introduced associated with the quark mass. The
only possibility is to promote the mass to an N=2 vector multiplet
associated with $U(1)_B$ \cite{N=2soft}.  Switching on its
auxilliary field vevs induce the tree level supersymmetry breaking 
operators
\beq
2 Re( F_{M}  q \tilde{q}) + D_{M} \left( |q|^2 -
|\tilde{q}|^2 \right)
\eeq
Again a number of consistency checks support the identification of these field
theory breakings with the angles between D4 branes \cite{nevans2}. There are
three independent real parameters in both the field theory and the brane
picture. The scalar masses in the field theory 
break $SU(2)_R$ but leave two  $U(1)_R$
symmetries of the supersymmetric theory intact. The scalar masses may 
always be
brought to diagonal form by an $SU(2)_R$ transformation that mixes $q$
and $\tilde{q}^*$. In the resulting basis there is an unbroken $U(1)$
subgroup of $SU(2)_R$.
In the brane picture the
D4$'$ branes lie at an angle in the $x^6-x^9$ directions breaking the
$SU(2)_R$ symmetry but leaving two $U(1)_R$ symmetries unbroken. }
\end{itemize}

The NS5 branes may be rotated in such a way as to preserve $N=1$
supersymmetry (corresponding to, for example, setting $F_2 = i F_1$ in
the field theory) and the resulting configuration with the adjoint
matter field decoupled is given by

\beq 
\begin{tabular}{|c|c|c|c|c|c|c|c|c|}
\hline
 & $\#$ & $R^4$ & $x^4$ & $x^5$ &  $x^6$ & $x^7$ & $x^8$ & $x^9$ \\
\hline 
NS5 & 1 & $-$ & $-$ & $-$  &  $\bullet$ & $\bullet$ & $\bullet$ & $\bullet$ \\
\hline
D4 & $N$ & $-$  & $\bullet$ &  $\bullet$ &  $[-]$ & $\bullet$ & 
                              $\bullet$ & $\bullet$ \\
\hline
NS5' & 1 & $-$ & $\bullet$ & $\bullet$ & $\bullet$ & $-$  &  $-$ & $\bullet$ \\
\hline
D4' & $F$ & $-$  & $\bullet$ &  $\bullet$ & $-$ & $\bullet$ & 
                              $\bullet$ & $\bullet$ \\
\hline
\end{tabular} 
\eeq

This configuration therefore describes an N=1 $SU(N)$ theory with $F$
matter fields. This is the theory that displays non-abelian
electro-magnetic duality and will be our starting point below.

\section{N=1 Duality}

The brane configuration of (4) describes N=1
SQCD. The duality of the field theory can be deduced from the brane
configuration as follows \cite{branedual}. 
Assuming the theory has an IR fixed point then
motion of the two NS5 branes relative to each other, which changes the
effective gauge coupling, should be a modulus of the theory. In that case
we can ask what happens when the two branes pass through each other. The
result has been deduced from string theory charge conservation and
winding number rules. The configuration is T dual to that discussed by
Hanany and Witten \cite{hanwit}; 
they showed that by conservation of magnetic monopole
number on the NS5 brane and equivalently by a linking number argument of
the bulk fields that the number of D4s ending on the right of the NS5
brane minus those
ending on the left must be conserved. Applying this rule to the brane
configuration of (4) we obtain, after passing the two NS5 branes
through each other, a configuration with $F-N$ D4 branes between the
two NS5 branes and $F$ semi-infinite branes. The new configuration
describes an $SU(F-N)$ gauge theory with $F$ flavours. This is Seiberg's
dual. A more quantitative approach is to move to M-theory on finite
compactified radius and perform these motions \cite{Mdual} 
which we review next.

The M-theory description of the N=1 SQCD configuration has an M5 brane
that is flat in $R^4$ and described in the remaining 7 dimensions (we
define $v = x^4 + i x^5$, $w = x^7 + i x^8$ and $t = exp(x^6+ix^{10}/R$)
by the curve
\beq v = z, \hspace{1cm} w = {\xi \over z}, \hspace{1cm} t =
z^{N}m^{F-N}/(z-m)^{F} \eeq
The curve has two $U(1)$ symmetries associated with rotations in the $v$
and $w$ planes. The parameters of the curve break the symmetries but may
be made to transform spuriously
\beq 
\begin{array}{c|ccccc}
& v & w & z & m & \xi\\
\hline
U(1)_v & 2 & 0 & 2 & 2 & 2 \\
U(1)_{w} & 0 & 2 & 0 & 0 & 2  \\
\end{array}
\eeq
These symmetries  correspond to
the field theory symmetries
\beq \label{v}
\begin{array}{c|ccccc}
& W & Q & \tilde{Q} & m & \Lambda^{b_0}\\
\hline
U(1)_R & 1 & 0 & 0 & 2 & 2(N-F) \\
U(1)_{R'} & 1 & 1 & 1 & 0 & 2 N \\
\end{array}
\eeq
where $\Lambda = exp(2\pi i \tau/b_0)$ and $b_0 = 3N-F$. We may
make the identifications $\xi = \Lambda^{b_0/N} m^{F/N}$ and $m$ is the
matter field mass.
The UV field theory displays $Z_N$ and $Z_{N-F}$ discrete subgroups of these  
symmetries. Viewing the curve asymptotically and as $\Lambda
\rightarrow 0$
\begin{eqnarray}
z \rightarrow \infty & w=0 & t = v^{N-F} m^{F-N} \nonumber\\
z \rightarrow 0 & v = 0 & t = \left( {1 \over w}\right)^N 
\Lambda^{b_0} m^{F-N}
\end{eqnarray}
The $U(1)_v$ and $U(1)_w$ symmetries (allowing $m$ to transform
spuriously but not $\Lambda$) are indeed asymptotically broken to 
$Z_N$ and $Z_{N-F}$ discrete subgroups. Other combinations of
the two $U(1)_R$ symmetries may also be identified in the asymptotic
curve. For example $U(1)_A$ symmetry is given by the rotations
\beq
\begin{array}{c|cccccc}
& v & w & t & z & \Lambda^{b_0} & m \\
\hline
U(1)_A & -2 & 2 & 0 & -2 & 2F & -2 \\
\end{array}
\eeq

The N=1 theory behaves like supersymmetric Yang Mills theory below the
matter field mass scale and has $N$ degenerate vacua associated with the
spontaneous breaking of the low energy $Z_{N}$ symmetry. In the curve this
corresponds to the $N$ curves in which $\xi_n = \xi_0 exp(2\pi
in/N)$ (equivalently $\Lambda^{b_0}_n = \Lambda^{b_0}_0 exp(2\pi
in)$). In the UV these curves can be made equivalent
by a $Z_{N}$ transformation.

For $F>N$ the curve may be recast in such a way that the dual picture of
the strong dynamics as $\Lambda \rightarrow \infty$ appears. In
particular we make the transformations
\begin{eqnarray}
m & \equiv & M^{N/F-N} \Lambda^{-b_0/F-N} \nonumber \\
\Lambda^{b_0} & \equiv & \tilde{\Lambda}^{-\tilde{b}_0} \nonumber \\
z & \equiv & {M^{F/F-N} \Lambda^{-b_0/F-N} \over z'}
\end{eqnarray}
and the curve is transformed to
\beq
v = {M^{F/\tilde{N}} \tilde{\Lambda}^{\tilde{b}_0/\tilde{N}} \over z'},
\hspace{1cm}
w = z', \hspace{1cm}
t = { M^{F-\tilde{N}} z^{'\tilde{N}} \over (M-z')^F}
\eeq
where $\tilde{N} = F-N$ and $\tilde{b}_0 = (3\tilde{N}-F)$.

In the limit $\tilde{\Lambda} \rightarrow 0$ at fixed $M$ this curve
degenerates to the IIA configuration describing a $SU(F-N)$ gauge theory
with $F$ quark flavours of mass $M$. This theory is 
Seiberg's dual SQCD theory. Note that in terms of the electric curve
this configuration is obtained when $\Lambda \rightarrow \infty$ with
$m$ scaled to zero appropriately to keep $M$ fixed. The duality of the
field theory is a strong-weak duality.

It is also of interest to note that the UV dual theory breaks
supersymmetry as observed in \cite{Mdual}. 
Including a mass term $m Q \tilde{Q}$ in
the electric theory maps in the magnetic variables to a superpotential
term $mM$ or equivalently to a shift in the vev of the auxilliary
component of $M$. The dual gauge dynamics restores supersymmetry in the
IR. In the brane picture this effect shows itself as follows: in the
electric variables the mass term is a relative displacement in the $w$
direction of the end point of the semi-infinite D4 brane and the $w$
coordinates of the $v$ plane NS5 brane. After the duality motion the
semi-infinite D4s connect directly to that NS5 brane and hence ly at an
angle in the $w$ direction. This is precisely the configuration we
identified in section 2 with corresponding to a shift in the vev of the
auxilliary field of the adjoint chiral multiplet of the N=2 flavour
gauge group, in other words, $M$. The duality maps interchanges an
electric theory mass with a magnetic theory $F_M$ term.

\section{N=0 Duality}

In section 2 we have reviewed how non-supersymmetric gauge theories may
be constructed in type IIA string theory. Adjoint matter fields may be
given masses by rotations of the NS5 branes and scalars by placing the
two sets of D4 branes at angles to each other. For the purposes of this
discussion we shall consider deformations from N=1 SQCD that restrict
rotations of the NS5 branes to those 
corresponding to a gaugino mass with the adjoint
chiral multiplet always decoupled. At the level of the string theory
these deformations of the brane configuration do not seem to preclude
the brane motions required to derive duality for the supersymmetric
theory. The relative angles of the branes do not disrupt the linking
number argument of Hanany and Witten. Naively we may move the branes
through each other as before and obtain an $SU(F-N)$ gauge theory with
$F$ matter flavours. Any rotations of the NS5 branes will presumably
remain and so a gaugino mass in one set of variables maps to a gaugino
mass in the other variables. Similarly an angle between the sets of D4s
in the $x^9$ direction will remain after the motion and thus a $D_M$
mass term in the electric variables maps to the same in electric
variables.

To place these observations on more solid and quantitative grounds we
will move to the M-theory curve describing such a brane configuration. A
curve describing a subset of possible configurations has been proposed
in \cite{nevans2}. The non-holomorphic curve 
\beq v = z +{\bar{\epsilon}\over \bar{z}} 
\hspace{1cm} w = {\Lambda^{b_0/N} m^{F/N}\over z}, \hspace{1cm} t =
z^{N}m^{F-N}/(z-m)^{F}, \hspace{1cm} x^9 = 4 \epsilon^{1/2} Re \ln z
\eeq
is a minimal area embedding ($\epsilon^{1/2}$ must be real for the
configuration to be single valued in $x^9$).

In the $R\rightarrow 0$ limit the
D4 branes lie in the $x^6$ and $x^9$ direction and the NS5
brane lying in the $w$ direction has been rotated in a non-supersymmetric
fashion into the $v$ direction.
We generically expect the field theory to have the 
supersymmetry breaking terms
\beq   
D \left( |q|^2 - |\tilde{q}|^2 \right) + m_\lambda \lambda \lambda
\eeq

We may identify the
parameter $\epsilon$ with field theory parameters from its symmetry
charges. Since it is chargeless under both $U(1)$s we may only identify it
as a function of
\beq
m_{\lambda}^N \Lambda^{b_0} m^F
\eeq

To complete the identification we note that the field theory retains a
$Z_F$ subgroup of $U(1)_A$ symmetry even after the inclusion of the soft
breaking terms. Requiring this property of the curve asymptotically
forces
\beq
\epsilon =  \left( m_{\lambda}^N \Lambda^{b_0} m^F \right) ^{1/N}
\eeq

Asymptotically the curve is then
\begin{eqnarray}
z \rightarrow \infty & w=0 & t = v^{N-F} m^{F-N} \nonumber\\
z \rightarrow 0 & v = {\bar{m}_{\lambda}  \bar{w}}  & 
t = \left( {1 \over \bar{v}}\right)^N 
\bar{m}_\lambda^N\Lambda^{b_0} m^{F-N}
\end{eqnarray}
which posseses a $Z_F$ subgroup of the U(1) 
\beq
\begin{array}{c|cccccccc}
& v & w & t & z & \Lambda^{b_0} & m & m_\lambda & D\\
\hline
U(1)_A & -2 & 2 & 0 & -2 & 2F & -2 & 0 & 0\\
\end{array}
\eeq

It is important to understand to what extent this identification of the
parameters between the brane configuration and the field theory is
valid. The identification has been made in the $R \rightarrow 0$ limit
where the field theory on the surface is perturbative. As $R$ is
increased or $\Lambda$ increased from zero we have no holomorphy
properties
and corrections between the brane parameters and field theory
which are arbitrary functions of the symmetry neutral quantities $|m|,
|\Lambda|$ and $|m_\lambda|$ may enter the relationship. The
perturbative identification is a helpful labelling of the parameters
because it displays their symmetry charges more readily. 

It is natural to ask what has become of the dual description of the N=1
theory? Again performing the redefinitions of the curve paramters using
(10) we obtain a curve for the dual variables
\beq
\begin{array}{c}
v = {M^{F/\tilde{N}} \tilde{\Lambda}^{\tilde{b}_0/\tilde{N}} \over z'} ~+~
\bar{m}_\lambda \bar{z}', \hspace{1cm}  w = z', \hspace{1cm}  
t = { M^{F-\tilde{N}} z^{'\tilde{N}} \over ( M-z')^F}\\
\\
x^9 = - 4 (m_\lambda M^{F/\tilde{N}}
\tilde{\Lambda}^{\tilde{b}_0/\tilde{N}})^{1/2} Re \ln z'
 
\end{array}
\eeq
The dual description (taking the $R \rightarrow 0$ limit) also has one
NS5 brane rotated and an angle between the semi-infinite and finite D4
branes in the $x^6$ $x^9$ plane. 
In the field theory the gaugino is massive and the dual squarks
are also massive through a non-zero D.

In fact we may take the decoupling limit for the gaugino mass, that is
take $\epsilon \rightarrow \infty$. $m$ is the only R-charged parameter
remaining and there is therefore nothing that can play the role of
either a gaugino mass or condensate fitting the assumption that the
gaugino has been decoupled.
We must define a new strong scale parameter below the gaugino mass
$\Sigma = m_\lambda^{N/F} \Lambda^{b_0/F}$ which has the R-charges of a
quark condensate and was identified as such in \cite{nevans2}. 
In fact we will see
below that in the light of the duality this precise identification must
be re-evaluated though the parameter is still plausibly related to a
quark condensate. 
The decoupled curve is
\beq \label{curve}
v= z + { (\bar{m} \bar{\Sigma})^{F/N}
\over \bar{z}}, \hspace{1cm} 
t = z^{N}m^{F-N}/(z-m)^{F},
\hspace{1cm} x^{9} = 4 (m \Sigma)^{F/2N} Re \ln z
\eeq

Again a duality map exists
\begin{eqnarray}
m & \equiv & M^{N/F-N} \Sigma^{-F/F-N} \nonumber\\
\Sigma & \equiv & \tilde{\Sigma}^{-1}\\
z' & \equiv & {m^{F/F-N} \Sigma^{-F/F-N} \over z} \nonumber
\end{eqnarray}
The dual curve is
\beq 
v= \bar{z}' + { (M \tilde{\Sigma})^{F/\tilde{N}}
\over \bar{z'}}, \hspace{1cm} 
t = z'^{\tilde{N}}M^{F-\tilde{N}}/(M-z')^{F},
\hspace{1cm} x^{9} = -4 (M \tilde{\Sigma})^{F/2\tilde{N}} Re \ln z' 
\eeq

Again the dual picture emerges from the M-theory curve in the limit
where $\Sigma \rightarrow \infty$ and $m \rightarrow 0$. Surprisingly
the string/M theory seems to be telling us that for massless quarks there
is a duality symmetry between an $SU(N)$ gauge theory with $F$ quarks
and an $SU(F-N)$ gauge theory with $F$ quarks. Presumably some components
of the dual meson of SQCD also survive though we have been
cavalier in the above discussion as to the boundary conditions at
infinity of the semi-infinite D4s. In the next section we address the
issue of whether there is any reasonable field theoretic 
interpretation of this duality that the curve has led us too.

\section{Duality In Non-Supersymmetric Field Theory}

Is the possibility that duality survives the breaking of supersymmetry
and the complete decoupling of the gaugino reasonable within the context
of the field theory? The first worry is that the massless fermions will
no longer satisfy the 'tHooft anomaly matching conditions in the
absence of the gaugino. In fact they do; the gaugino's mass breaks the
$U(1)_R$ symmetries of the supersymmetric model and the only remaining
symmetries are the flavour symmetries of the matter fields. Since the
gaugino does not transform under the flavour symmetries the other
fermions' anomaly matching for the flavour symmetries is maintained 
\cite{soft3}. 

Naively we would expect that with the decoupling of the gaugino there
would no longer be supersymmetry to stabilize the scalar fields masses
against radiative effects of order $m_\lambda$. If this were the case
then the low energy behaviour of the theory on the branes surface would
be that of non-supersymmetric QCD. Could the duality be one of QCD? We
must conclude that it could not. For the duality to link an $SU(N)$
theory to a flavour dependent $SU(F-N)$ theory requires that for the
mass branch (giving a mass to the quarks) 
of each of the two descriptions there is a higgs branch in
the other. QCD has no higgs branch and so there is no way to flow to an
$SU(N-1)$ gauge theory as one dual quark is given a mass. One might be
tempted to seek a description with four-Fermi interactions triggering a
higgs branch by a colour superconductivity mechanism (for example the
dynamical generation of a $\langle \psi_q \psi_M \rangle$
condensate). Such a mechanism is though limited by the strong belief
from NJL type models that a four Fermi or other higher dimension
operator has a none zero critical coupling for the triggering of chiral
symmetry breakdown. For the duality the higgs
branch would have to switch on for an infinitessimal breaking of the
flavour symmetry so the theory behaved correctly below the scale of the
quark mass.

It appears therefore that the brane picture above is
not decoupling the scalar fields and they remain light or at least
relevant to the low energy dynamics. This is presumably an artifact of
moving to strong coupling. The duality is observed in the brane picture
when $\Lambda$ is taken to infinity, that is the strong coupling scale
is taken to the UV cut off. Tree level massless scalar fields are not in this
limit given the opportunity to radiatively acquire masses above the
strong coupling scale and decouple. Alternatively we must bare in mind
that the curve may not be describing the global minimum of the
theory. There may have been a first order phase transition for some
value of the supersymmetry breaking and the global minimum is described
by a curve other than (12). We would then be studying a local minimum of
the theory where the scalars might have vevs at the expense of a large
cost in energy. Nevertheless the continued existence of duality at this
local minimum is interesting. The correct question therefore is
whether non-supersymmetric field  theories with fine tuned 
light scalars can
continue to exhibit duality. Once we include the scalar fields the
correct flow in the two theories with the addition of mass terms (which
become $F_m$ terms in the dual) is correctly satisfied following the
SQCD pattern and the duality is possible.

We must also ask whether the duality continues to make sense in the
presence of a $D_M$ term from (3). Including such a term in the electric
variables triggers a higgs branch through the potential term
\beq
(D_M + |Q|^2 - |\tilde{Q}|^2)^2
\eeq
In the brane picture this can be seen because when the semi-infinite D4
branes are placed at an angle into the $x^9$ direction the D4 branes may
decrease their lengths and hence tension by detaching from the central
NS5 brane. They are then free to move along the single NS5 brane to which
they are attached. The result is that the gauge group is higgsed of one colour
and one flavour of matter is lost. In the dual construction the
equivalent motion of the semi-infinite D4s along that NS5 brane now does
not involve the $F-N$ D4s associated with the gauge group. The motion
corresponds to a meson vev and hence a mass branch in the dual. In the
field theory this is provided by the potential
\beq
(D_M + |q|^2 - |\tilde{q}|^2 - [M^\dagger,M])^2
\eeq
which has a mass branch. The duality again is at least potentially
consistent with the addition of these soft breaking terms. It was the
assumption that this was the case in the supersymmetric limit that was
used in \cite{branedual} 
to derive the field theory duality by moving the NS5 branes
past each other in the $x^9$ direction.

As the scalars remain light in the theory the precise identification of
the parameter $\Sigma$ is harder. 
Perturbatively softly breaking the supersymmetric field theory with a
gaugino mass is known to generate a quark condensate since
\beq
\langle \psi_Q \psi_{\tilde{Q}} \rangle \simeq F_M \simeq m_\lambda m^{N/F-N}
\Lambda^{b_0/N} 
\eeq
In \cite{nevans2} where it was assumed that the scalars radiatively decoupled
$\Sigma = m_\lambda^{N/F} \Lambda^{b_0/F}$ was identified with the quark
condensate in the decoupling limit when $m_\lambda \gg m$.
If the scalars remain light and the theory possesses a strong weak duality
then presumably at the origin of moduli space of the massless theory the
theory is conformal. This can be seen since by suitable
choice of large $N$ and $F$ we may place the massless
theory (with the gaugino decoupled) arbitrarily close to
where asymptotic freedom is lost and where the theory is believed to
have a weakly coupled IR Banks-Zaks fixed point. At the origin of moduli
space, if the duality is
correct in the field theory, then there is a strongly interacting theory
at a conformal fixed point that flows to the same low energy physics. 
These theories can not dynamically generate a quark
condensate since they're conformal
and so we conclude that a condensate only develops with the
introduction of a mass term. 

In the decoupling limit to bring the M-theory curve to a standard form
we rescaled the curve parameter $z$ by $m_\lambda$. In the field theory
this corresponds to rescaling the dual fields by $m_\lambda$ and this is
presumably the correct basis for the decoupled field theory. In this basis
the vev of the scalar, $a_M$, has the same symmetry
charges as the quark
condensate. This is why the condensate appears as the mass parameter in
the curve (24). In the field theory we expect the scalar vev and quark
condensate to grow together. 

The possibility then is that non-supersymmetric
$SU(N)$ gauge theories with $F$ quark and scalar fields 
(fine tuned to masslessness) in the
fundamental representation is dual to an $SU(F-N)$ gauge theory with $F$
dual quarks and scalars plus $F^2$ singlet meson fermions and
scalars. In the massless limit the dual theories do not generate a
fermion condensate but as a mass is introduced a fermion condensate and
scalar vev develop together and the theory has a mass gap. 
Of course  we emphasise again that the brane construction strong
coupling expansion  does
not prove this for these field theories though they are 
amusingly suggestive.

The failure of the scalar fields to decouple even in the absence of
supersymmetry is awkward for the discussion of true QCD
dynamics. Ideally we would like to decouple the scalars at tree level
and not rely on radiative effects. Unfortunately the deformations of the
brane configuration do not appear to correspond to a suitable scalar mass
term. The best we could hope to achieve is to switch on masses of the
form of (3) but these masses are always unbounded and trigger a higgs
branch of the theory. We do not know of anyway to give all scalars a
positive tree level mass and so their decoupling remains frustratingly
elusive. It is though clear that a flavour dependent dual gauge theory
such as the branes suggest
is not possible for QCD since it possesses no higgs branch. \vspace{2cm}

\noindent {\bf Acknowledgements:} This work was in part supported by the
Department of Energy under contract $\#$DE-FG02-91ER40676. \vspace{2cm}



\end{document}